# Hawking- Unruh Radiation from the relics of the cosmic quark hadron phase transition

Bikash Sinha

Variable Energy Cyclotron Centre
1/AF Bidhan Nagar, Kolkata 700064, India
Email: bikash@vecc.gov.in; bsinha1945@gmail.com

**Abstract**:

It is entirely plausible that during the primordial quark – hadron transition, microseconds after the Big Bang, the universe may experience supercooling accompanied by mini inflation leading to a first – order phase transition from quarks to hadrons. The relics, in the form of quark nuggets expected to consist of Strange Quark Matter, with a baryon number beyond a critical value will survive.

It is conjectured that color confinement turns the physical vacuum to an event horizon for quarks and gluons. The horizon can be crossed only by quantum tunnelling. The process just mentioned is the QCD counterpart of Hawking radiation from gravitational black holes. Thus, when the Hawking temperature of the quark nuggets gets turned off, tunnelling will stop and the nuggets will survive forever. The baryon number and the mass of these nuggets are derived using this theoretical format. The results agree well with the prediction using other phenomenological models. Further, the variation of Hawking temperature as a function of baryon number and mass of the nugget mimicks chiral phase transition, somewhat similar to the QCD phase transition just described. Finally the strange quark nuggets may well be the candidates of baryonic dark matter.

_________________________________________________________________________

Microseconds after the Big Bang the universe consisted of quarks, leptons and photons. As the universe expanded and cooled to a temperature around (150-200) MeV the chirally symmetric quark gluon plasma made a phase transition to the hadron phase of broken chiral symmetry. For very small baryon number η = $n_B$/s ~ $10^{-9}$; ($n_B$, the net baryon density and s the entropy density), the wisdom of lattice suggests a rapid "cross- over" from the universe of quarks to the universe of hadrons. A first- order QCD phase transition seems unlikely and any imprint of the time before the phase transition gets erased out. However, lattice calculations for an expanding universe with massless bare quarks is not entirely satisfactory. Instead, chiral models of QCD have been used extensively [1,2,3,4].

It is conventionally assumed that the baryon asymmetry $\eta = (n_B - n_{\bar{B}})\gamma$ at that primordial epoch of phase transition is the same as that of today's universe $\eta \sim 10^{-10}$. There are however reasonably straightforward arguments [1,2,3,4,5,6,7] that η at that epoch is much higher and indeed of the order of $\eta \sim \mathcal{O}(1)$ unity. However, it is seen that after the phase transition η goes back to $10^{-10}$, as it is today. The consequence of such a possibility is discussed in the following.

Witten [7] and others [4,5,6,8] have argued that a first order phase transition is plausible with a "small" supercooling. In a recent private communication Witten [5] further asserted that if $\eta \approx n_B/\gamma$ remains $\sim 10^{-10}$ at the point of q-h phase transition, as it is in the current universe, then supercooling is implausible. However, he also points out [5] that if the baryon to photon ratio is not small during the QCD phase transition and becomes small because of some phenomena at later times, then supercooling is plausible in principle. In the following, we demonstrate that this is entirely possible.

This is the central issue, the relevance of baryon asymmetry at that primordial epoch.

One of the more compelling scenarios of baryogenesis is based on its generation from leptogenesis through topological sphaleron transitions occurring around the electroweak transition temperature. Leptogenesis occurs through out-of-equilibrium decays of heavy righthanded neutrinos which occur naturally within a seesaw mechanism, leading to Majorana masses for neutrinos, (as well as neutrino oscillation parameters) within observable ranges. Fermions with only Majorana masses and no Dirac masses ( Majorana fermions) are charge self-conjugate spin- 1/2 particles for any global U(1) charge. If this U(1) charge is associated with lepton number, then the charge self-conjugate property automatically implies that Majorana mass terms violate lepton number. Thus, it is this supposedly Majorana nature of neutrinos (even if they have a Dirac component as well) which lies at the heart of the incipient lepton number violation. The positive aspect of this mechanism of leptogenesis-induced baryogenesis is that one obtains a numerical result close to the observed baryon-photon ratio of $\mathcal{O}(10^{-10} - 10^{-9})$ without any fine tuning.

The resolution of the issue as to whether neutrinos are predominantly Majorana fermions, as happens to be the common prejudice currently, is to be decided by the currently ongoing experiments on neutrinoless double beta decay. If, contrary to extant belief, such experiments happen to yield null results, and neutrinos are confirmed to be Dirac fermions, this scenario of baryogenesis loses its prime attraction, entailing unsavoury fine tuning.

Given such a volatile situation, alternative scenarios of baryogenesis cannot be ruled out. Prominent among these is the nonthermal Affleck-Dine mechanism [13].

The Affleck-Dine mechanism [13] has the potential to produce a baryon asymmetry of $\mathcal{O}(1)$ without requiring superhigh temperatures. However, the observed baryon asymmetry of $(10^{-10})$ at CMB temperatures needs to emerge naturally from such a scenario. This is what is achieved through a "little inflation" of about 7 e-folding occurring at a lower temperature, which may be identified with the QCD first order phase transition [1,3]. Such an inflation naturally dilutes the baryon photon ratio to the observed range, even though the baryon potential before the first order phase transition may have been high (of $\mathcal{O}(1)$ in photon units). Comparing this "little inflation" with the more standard Guth's inflationary model [14], one finds that the patterns of entropy variation in the two cases are very different. In the standard inflationary model [14] the entropy is conserved during exponential expansion, and increases, due to reheating when bubbles collide, at the end of the transition. However in Guth's scenario, supercooling is there though very large; In the little inflation scenario for the case of quark-hadron phase transition, on the other hand, the entropy is constantly increasing during the quark-hadron phase transition.

Bhattacharya et al. [9] however, have demonstrated, using Flux Tube model that quark nuggets of baryon number $\sim 10^{43}$, or larger will survive even upto now. These nuggets will consist of Strange Quarks (SQN) [7] and could be a viable candidates of cold dark matter [5,7,10].

We shall now discuss the evolution of these SQNs as the universe expands with time.

The SQNs remain in kinematic equilibrium due to radiation pressure (photons and neutrinos) acting on them with their velocity extremely non-relativistic. Effectively, the SQNs are almost static. It has been argued [10] that when the temperature of the universe is around couple of MeV, gravity dominates over radiation and the SQNs grow in size after collapsing under gravitational attraction, due to other SQN's; it can not grow however beyond a certain mass [10]. For baryon number $10^{42}$ the mass of the collapsed SQN is $0.24 M_\odot$ but for baryon number $10^{46}$ it goes down to $0.0001 M_\odot$ [10]. These inputs are crucial for the construction of the model developed in the following.

Color confinement in QCD does not allow colored constituents to exist in the physical vacuum and thus in some sense color confinement is similar to gravitational confinement provided by black holes. Thus, while 'G' ensures gravitational attraction and eventually to black holes, 'B', the Bag pressure ensures confinement of quarks leading to colorless physical vacuum (white holes) in QCD.

It has been argued [11] that quantum tunnelling through a color event horizon is the QCD counterpart of Hawking- Unruh radiation from the gravitational black holes.

Let us consider the SQNs, after the gravitational collapse as QCD white holes, with neutrons tunnelling out from the event horizon. The natural length scale for SQNs, heuristically, can be argued as [8] $L_B = M_N/M_\odot (B^{1/4})^{-1}$, $M_N$ being the mass of the SQN and B the Bag Pressure which keeps the color confined. So, the entropy for QCD white hole becomes

$$S_{QN} = (A/4L_B^2) = \pi R^2 B^{1/2} \left(M_N/M_\odot\right)^{-2} \qquad (1)$$

The celebrated Hawking entropy can be written as

S=$\pi R^2$/G; for QCD [11], G≡1/2σ; σ=0.16 GeV$^2$, σ being the string tension.

The confinement radius R of a SQN is equivalent to the "strong" Schwartzchild radius equivalent of the gravitational black hole.

$$S = 2\pi R^2 \sigma = \pi R^2 B^{1/2} (M_N/M_\odot)^{-2}, \qquad (2a)$$

thus $B^{1/2} \left(M_N/M_\odot\right)^2 = 2\sigma$, it is noted that in the final expression, (2b)

"strong" schwartzchild radius gets crossed out as it should. For $B^{1/4}$=150 MeV; $(M_N/M_\odot)$=0.265 and for $B^{1/4}$=200 MeV $(M_N/M_\odot)$=0.35; the result agrees closely with the results obtained by Banerjee et al., [10], although derived from a very different route of arguments.

Using the conjectured analogy between black hole thermodynamics and the thermodynamics of confined charged quarks, the SQNs, we transliterate black hole mass, charge and gravitational constant [11] into the mass $M_N$ of the SQNs, baryon number $B_N$ and the string tension $\sigma$

$$\{M, Q, G\}_{BH} \leftrightarrow \{M_N, B_N, 1/2\sigma\}_{QN}$$

such that

$$T_{QN}(B_N, M_N) = T_{QN}(B_N, M_N = 0) \times \left[\frac{4\sqrt{1 - M_N^2/2\sigma B_N^2}}{\left(1 + \sqrt{1 - M_N^2/2\sigma B_N^2}\right)^2}\right] \quad (3)$$

In Fig 1, $T_{SN}(B_N, M_N)/T_{SN}(B_N, M_N = 0)$ as a function of $(M_N^2/2\sigma B_N^2)$ is shown [11]; the mass of $M_N$ being zero seems to imply complete restoration of chiral symmetry, similarly when $M_N = \sqrt{(2\sigma)}B_N \approx B_N$ all the baryons (quarks) have a total mass approximately the same as Baryon number, chiral symmetry is exactly and completely broken. For M ~ $10^{44}$ (GeV) [10] $B_N = (1/\sqrt{(2\sigma)}M_N) \approx 6 \times 10^{43}$; Hawking radiation stops and no more tunneling out to the physical vacuum takes place. The SQNs survive the evolution of the universe. The surviving nugget with baryon number ~ $6 \times 10^{43}$ agrees closely the value obtained by Bhattacharya et al. [9], although obtained from a flux tube model.

The phase transition is not instantaneous but happens over a period of time [12]; for a critical temperature of transition 100 MeV, the time it takes for the phase transition ≈ $t_{ch}$ ~ 144 μsec [12] and for a critical temperature 150 MeV $t_{ch} \equiv 64$ μsec; during this time all the quarks inside the nugget, acquires mass, by breaking chiral symmetry.

One has attempted so far utilizing the possible equivalence of gravitational black hole and QCD white hole to drive the mass of QCD white hole which survives the evolution of the universe uptil now. Using the same conceptual framework we have derived the baryon number at which evaporation or equivalently tunnelling out of the QCD white hole stops. Stopping of Hawking radiation from white holes indicate the survivability of the nuggets. The results agree closely with previously obtained results [9, 10] using somewhat different theoretical framework. The results obtained using black hole analogy is quite straightforward but based on fundamental physical laws, such as thermodynamics and Hawking radiation from black hole. The results obtained using black (white) hole thermodynamics lends credence to the framework used and the results obtained.

Where do these SQNs go? Originally, it was proposed that these SQNs are the MACHOs, the Massive Astrophysical Compact Objects [10], detected in the direction of the Large Magellanic Cloud (LMC) of mass range $(0.15\text{-}0.95)M_\odot$, with a probable mass of $0.5M_\odot$, and the total number being $N_{macho} \approx 10^{23\text{-}24}$. These MACHOs are candidate of baryonic dark matter [10].

**Acknowledgement:**

The author wishes to thank E. Witten for his perceptive comments on supercooling. He is grateful to Dima Kharzeev, Larry Mc Larren and Roger Penrose for their suggestions. Comments from Pijus Bhattacharya, Debasish Mazumdar and Sibaji Raha are gratefully acknowledged. Finally all the help rendered by Dr. Chiranjib Barman is appreciated.

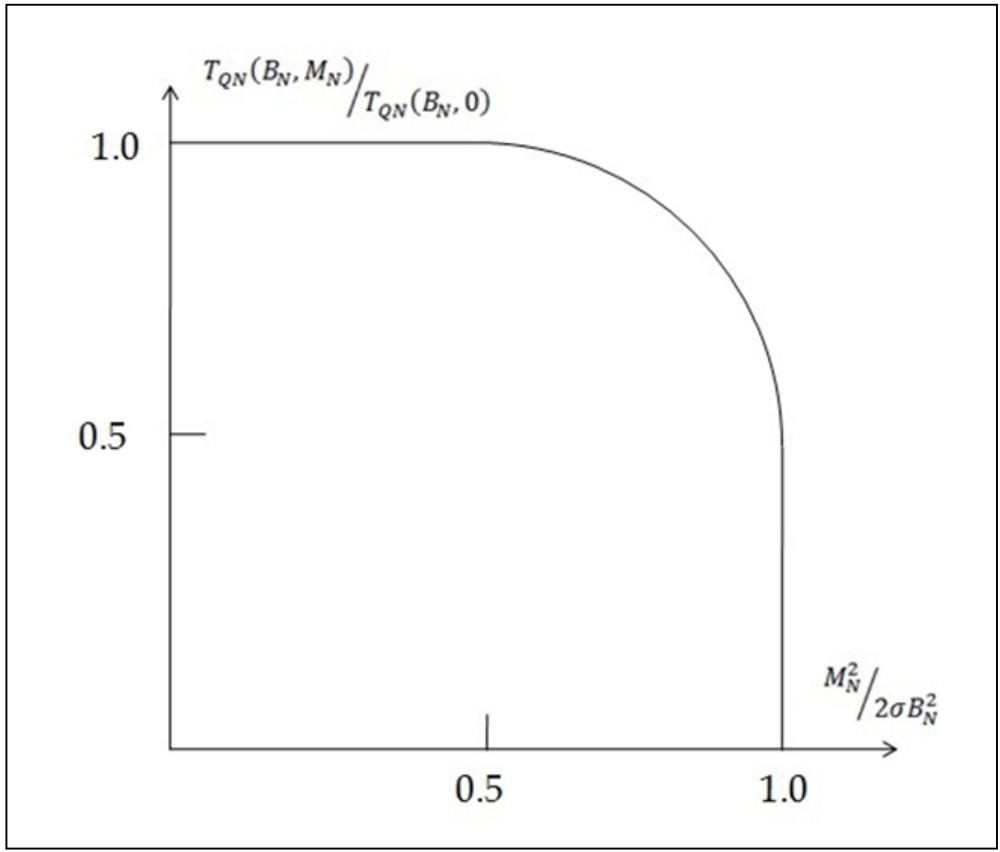

**Fig 1**: Hawking temperature as a function of mass $M_N$ and the baryon number $B_N$ of the nuggets; σ is the string tension.